  \documentclass[useAMS,usenatbib, letterpaper]{mn2e}

\usepackage[a4paper]{geometry}
\usepackage{multicol}
\usepackage{lscape}
\usepackage{rotating}

\usepackage{graphicx}
\usepackage{subfigure}
\usepackage{color}
\usepackage[authoryear]{natbib}
\usepackage{names}

\def\eg{{\it e.g.} }

\newcommand\blfootnote[1]{%
  \begingroup
  \renewcommand\thefootnote{}\footnote{#1}%
  \addtocounter{footnote}{-1}%
  \endgroup
}

\def\Herschel{{\it Herschel } }
\def\Herscheldot{{\it Herschel. } }
\def\Herschelbelong{{\it Herschel's } }
\def\Herschelnospace{{\it Herschel} }

\def\kcrb{ $\kappa$ CrB }
\def\kcrbcomma{ $\kappa$ CrB, }
\begin{document}
\title[]{Herschel Observations of Debris Discs Orbiting Planet-hosting Subgiants}

  \author[A. Bonsor et al.]{Amy Bonsor$^{1,2}$\thanks{Email:amy.bonsor@gmail.com}, Grant M. Kennedy$^3$, Mark C. Wyatt$^3$, \newauthor John A. Johnson$^4$ and Bruce Sibthorpe$^5$ \\
$^1$UJF-Grenoble 1 / CNRS-INSU, Institut de Planétologie et d'Astrophysique de Grenoble (IPAG) UMR 5274, Grenoble, F-38041, France  \\     
$^2$School of Physics, H. H. Wills Physics Laboratory, University of Bristol, Tyndall Avenue, Bristol BS8 1TL, UK \\
$^3$Institute of Astronomy, University of Cambridge, Madingley Road, Cambridge CB3 OHA, UK \\
$^4$Harvard-Smithsonian Center for Astrophysics, 60 Garden Street, Cambridge, MA 02138, USA \\
$^5$SRON Netherlands Institute for Space Research, Zernike Building, P.O. Box 800, 9700 AV Groningen, The Netherlands \\
}

\maketitle

\begin{abstract}

Debris discs are commonly detected orbiting main-sequence stars, yet little is known
regarding their fate as the star evolves to become a giant. Recent observations of radial
velocity detected planets orbiting giant stars highlight this population and its
importance for probing, for example, the population of planetary systems orbiting
intermediate mass stars. Our  \Herschelnospace$^*$ survey observed a subset of the Johnson et al
program subgiants, finding that 4/36 exhibit excess emission thought to indicate debris,
of which 3/19 are planet-hosting stars and 1/17 are stars with no current planet
detections. Given the small numbers involved, there is no evidence that the disc
detection rate around stars with planets is different to that around stars without
planets. Our detections provide a clear indication that large quantities of dusty
material can survive the stars' main-sequence lifetime and be detected on the subgiant
branch, with important implications for the evolution of planetary systems and
observations of polluted or dusty white dwarfs. Our detection rates also provide an
important constraint that can be included in models of debris disc evolution.

\end{abstract}

\section{Introduction}
\label{sec:intro}

Belts\blfootnote{$^*$ \Herschel is an ESA space observatory with science instruments
provided by European-led Principal Investigator consortia and with important
participation by NASA } of rocks and dust, known as debris discs are commonly detected around main-sequence stars. Recent surveys found infra-red (IR) excess emission, a good indicator for the presence of a debris disc, around 15\% of FGK stars  and 32\% of A stars \citep{Beichman2006, Bryden06, Moromartin07, Hillenbrand2008,Trilling08, Greaves2009, su06}. The observed emission must result from small dust grains \citep[e.g.][]{wyattreview}, yet, theoretical estimates find that the lifetime of such small grains against both collisions and radiative forces is short. The collisional evolution of a population of larger parent bodies are, therefore, generally invoked to explain the observed debris discs \citep{wyattreview}. Such collisional evolution naturally explains the decay with age in the fractional luminosity of observed debris discs \citep[\eg][]{Currie2008, su06,rieke05}.

As the number of resolved images of debris discs grow \citep[\eg][]{resolveddiskFomalhautbetapic, fomalahautresolvedscatteredlight, etatel, Churcher10}, so does the diversity of structures observed. In many cases, interactions with planets are invoked to explain the observations \citep[\eg][]{Churcher11, AugereauAUMic, Augereau01, chiang_fom}. It may be that the presence of debris discs correlates with the presence of planets, but there is as yet no strong evidence either way \citep{Bryden09, Moromartin07, Kospal2009, Wyatt61Vir}. 
Theoretically a correlation may be anticipated because the properties of the protoplanetary disc affect the outcome both for the debris disc and the planets \citep{WyattClarke07}, or  because dynamical effects such as instabilities in the planetary system can also have a significant effect on the debris disc \citep{Raymond2007}. This motivates observations of planets and debris discs orbiting the same stars.

The majority of confirmed planets are currently detected using the radial velocity (RV) technique\footnote{exoplanet.eu}. Radial velocity detections of planets orbiting main-sequence A stars are hindered due to high jitter levels and rotationally broadened absorption lines \citep{Galland05, Lagrange09}. Thus, previous efforts to compare the populations of planets and debris discs have focused on sun-like stars and RV planets \citep{Bryden09,Kospal2009, Moromartin07}. There are now a growing number of detections of planets around `retired' A stars, now on the subgiant or giant branch \citep[\eg][]{Johnson06,Johnson07,Bowler2010, Sato2010}, although some controversy does exist regarding the exact evolutionary paths of these stars \citep{Lloyd2011, Lloyd2013, Schlaufman2013}. These observations provide some key insights into the potential differences between the planetary population around sun-like and intermediate mass stars, that otherwise can only be probed by direct imaging of planets around main-sequence A stars \citep[\eg][]{hr8799detection08, fomb2008, Rameau2013}. Very little, however, is known regarding the population of debris discs around subgiants. By studying debris discs orbiting these `retired' A stars, we access a new population of planetary systems, from which more can be learnt regarding the structure and links between planets and debris discs, with a focus on intermediate mass stars.

In addition to providing a new and interesting sub-set of planetary systems to study, debris discs orbiting subgiants also provide evidence regarding the first step in the evolution of debris discs beyond the main-sequence. The interest in the fate of debris discs has grown with the growing evidence for planetary systems orbiting white dwarfs. Observations of both polluted \citep[\eg][]{ZK10, gaensicke2012, Koester11} and dusty \citep[\eg][]{farihi09,Barber2012} or gaseous \citep[\eg][]{Gaensicke06,Gaensicke2007, melis10} material very close to white dwarfs are thought to be linked with the presence of planets and planetesimal belts \citep[\eg][]{jurasmallasteroid, DebesSigurdsson, bonsor11, debesasteroidbelt}. For this to be true, both planets and debris discs must survive the star's evolution from the main-sequence to the white dwarf phase. The subgiant branch is the first step on this evolutionary path.

In this work we present \Herschel observations of a sample of 36 subgiants in which we search for excess emission, indicative of a debris disc. Radial velocity planets have been detected for half of the sample as part of the RV survey to search for planets orbiting `retired' A stars using Lick and Keck observatories \citep{Johnson06}, whilst the other half of the sample was observed as part of the same program, but no planets were detected. We start in \S\ref{sec:observations} by discussing our \Herschel observing strategy. This is followed in \S\ref{sec:results} by the results of our observations and a discussion of their meaning in \S\ref{sec:model} and \S\ref{sec:conclusions}.


\begin{figure}
\includegraphics[width=0.48\textwidth]{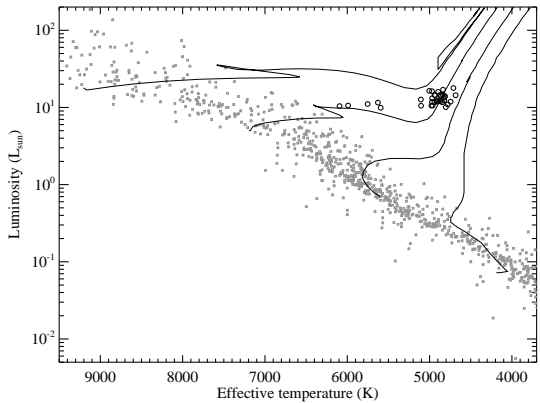}

\caption{Our sample plotted on a HR diagram (large black circles), alongside stars from the UNS sample (small grey dots), an unbiased sample of nearby main-sequence stars \citep{Phillips09} and stellar evolution tracks taken from \citet{Girardi2002} for 0.6, 1, 1.5, and 2.0 $M_\odot$. Tracks start at the zero-age main-sequence and move to the upper right over time. Our sample of subgiants have similar luminosities to main-sequence A stars, but are cooler. } 
\label{fig:hr}
\end{figure}

\section{Observations}
\label{sec:observations}
Observations were performed using the Herschel Photodetector and Array
Camera \& Spectrometer (PACS, \cite{Poglitsch2010}) at 100 and
160$\mu$m, as listed in Table 1.  These observations were performed in
mini scan-map mode with two observations being performed with a $40\deg$
cross-linking angle. Four repeats were used for each observation and
with eight scan legs per repeat.  The total observing time was
approximately 1790s per target. 

Data were reduced with the Herschel Interactive Processing Environment
version 7.0 Build 1931 (HIPE, \cite{Ott2010}) using version 32 of the PACS
calibration.  Some data from the telescope turn-around phase (when
scanning above $5\arcsec/s$) were used to minimize the ultimate noise
level.  Maps were then made using the HIPE photProject task to provide
`drizzle' maps \citep{Fruchter_Hook2002} with pixel scales of 1 and 2
arcsec in the 100 and 160\,$\mu$m bands respectively.  The data were
high-pass filtered to mitigate low frequency 1/$f$ noise, using
filtering scales of 66 and 102 arcsec (equivalent to a filter radius
of 16 and 25 PACS frames) in the 100 and 160\,$\mu$m bands
respectively.

The PACS point-spread function (PSF or beam) includes significant
power on large scales (10\% beyond 1 arcmin).  Consequently, the
filtering performed during the data reduction will reduce the flux
density of a source by $10-20\%$, due to the filter removing the ‘wings’
of the PSF.  For point sources this can be readily accounted for using
correction factors, determined from comparison of bright stars with
known fluxes with the PACS aperture flux. Correction factors of $1.19 \pm 0.05$ and $1.21 \pm 0.05$ at 100 and 160$\mu$m were determined from analysis of the DEBRIS (Disc Emission via a Bias-free Reconnaissance
in the Infrared/Submillimetre) survey \citep[\eg][]{Matthews2010} targets \citep{Kennedy99Her}. This can also be applied to resolved sources when the
source remains similar in scale to the beam Full Width Half Maximum (FWHM).

\subsection{Sample}
\label{sec:sample}
We consider here a sample of 36 stars, 19 of which have planet detections from \cite{Johnson10_subgiant, Johnson07, Johnson08, Johnson10, Johnson2011,Johnson2011_24Sex, Bryaninprep} and 17 of which are control stars, that have been searched for planets as part of a survey using Lick and Keck observatories to search for planets orbiting `retired' A stars \citep{Johnson06}, but where nothing was detected\footnote{The two samples were originally the same size, but planets have recently been found orbiting one of the proposed control sample.}.
 All stars were observed in the radial velocity programs such that the spectrum receives the same signal to noise ratio (S/N), regardless of stellar properties or sky conditions. Thus, the survey is complete to planets with velocity semiamplitudes $K> 20ms^{-1}$ and periods equal to the survey baseline of 6 years \citep{Johnson_planetpopulation}. The non-detection limits equate approximately to an absence of planets on orbits shorter than 200days, with a mass limit of around a Neptune mass at this semi-major axis, although the exact limits vary from target to target. The control sample were approximately matched to the planet sample in terms of distribution of stellar luminosity and distance. We only include stars within 160pc of the Sun, of mass greater than $1.5M_\odot$ and $L>15L_\odot$. This minimises the probability of their lying outside the local bubble where interactions with the ISM can mimic debris disc emission \citep{Kalas2002, Gaspar2008}, as well as maximises the probability of disc detection. The position of the sample on a HR diagram is shown in Fig.~\ref{fig:hr}.

\section{Results}
\label{sec:results}
\subsection{Photometry}

\label{sec:pointsource}
In order to analyse the sample for emission from debris discs, it is first necessary to account for the stellar contribution to the emission. Optical and near-infrared photometry is collected from numerous catalogues \citep{Morel1978, Moshir1993, Hauck1997, Perryman1997, Hog2000, Cutri2003, Mermilliod1987, Ishihara2010}. These data were used to find the best fitting stellar model, using the PHOENIX Gaia grid \citep{Brott2005}, via a $\chi^2$ minimisation, as in \cite{Kennedy99Her, Kennedy_binary, Wyatt61Vir}. This method uses synthetic photometry over known bandpasses and has been validated against high S/N MIPS 24$\mu$m data for DEBRIS targets, showing that the photospheric fluxes are accurate to a few percent for main-sequence, AFG-type, stars.

Most stars in this sample are faint and have predicted photospheric fluxes lower than the PACS detection limit, thus, any detected emission is likely to result from a debris disc, particularly at $160\mu$m. However, emission from background objects may contribute significantly to emission in the far-IR and sub-mm and without the stellar emission to guide the pointing, we are reliant on \Herschelbelong pointing to indicate the location of the star, which on average is accurate to within $1.32\arcsec$ in our observations \footnote{A weighted average of the absolute pointing error for obsID 1342216480-1342221945 = 2.36", 1342223464-1342225498 = 1.45", 1342229965-1342237471 = 1.1", 1342241949-1342243725 = 0.8"}. There is clear evidence that some previous \Herschel observations of debris discs have been contaminated by emission from background objects \citep[\eg][]{Donaldson2012}, particularly at $160\mu$m, although this is only found to be important for 1 or 2 of our sample (see \S\ref{sec:background}).

\begin{figure}
\includegraphics[width=0.48\textwidth]{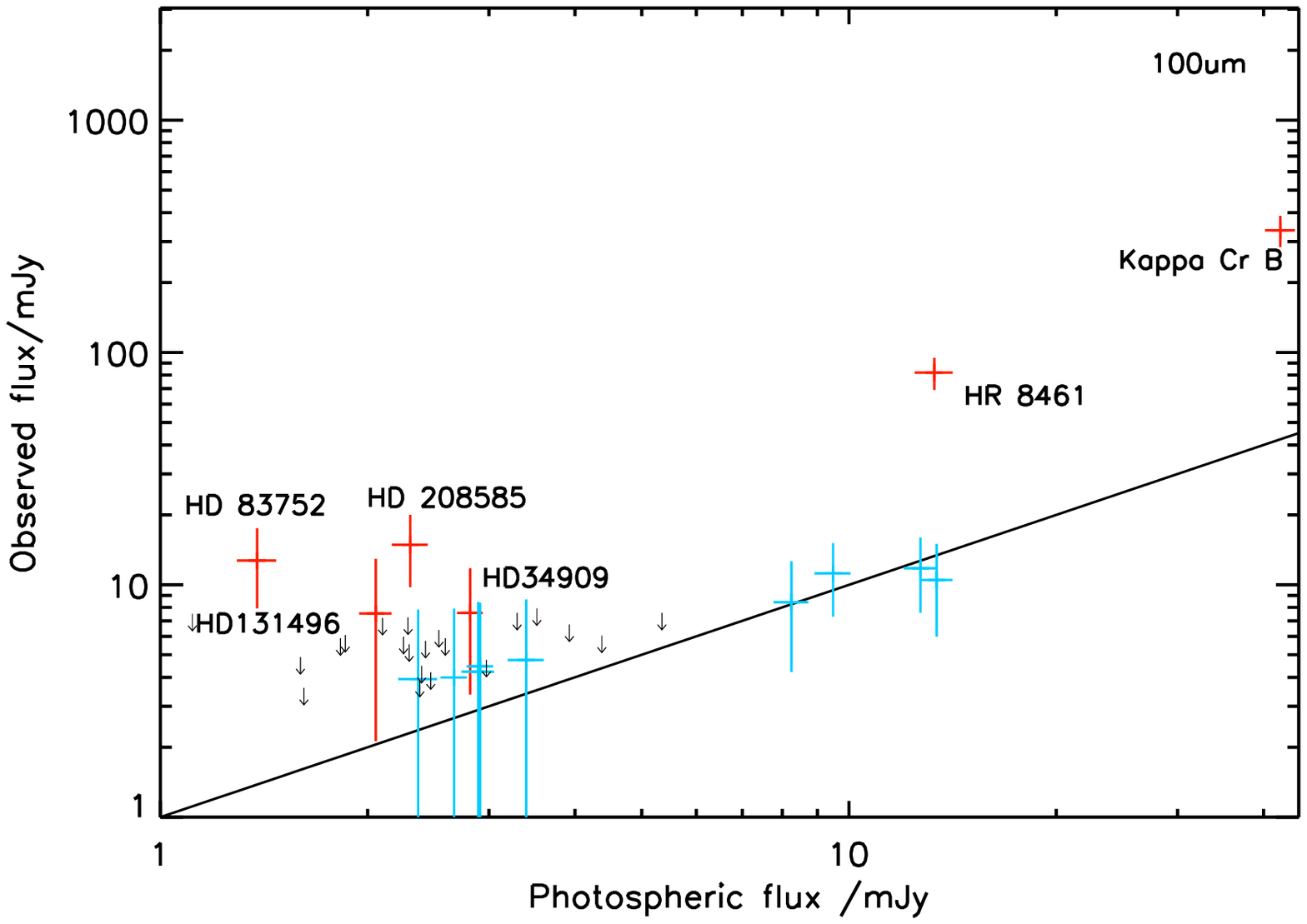}
\includegraphics[width=0.48\textwidth]{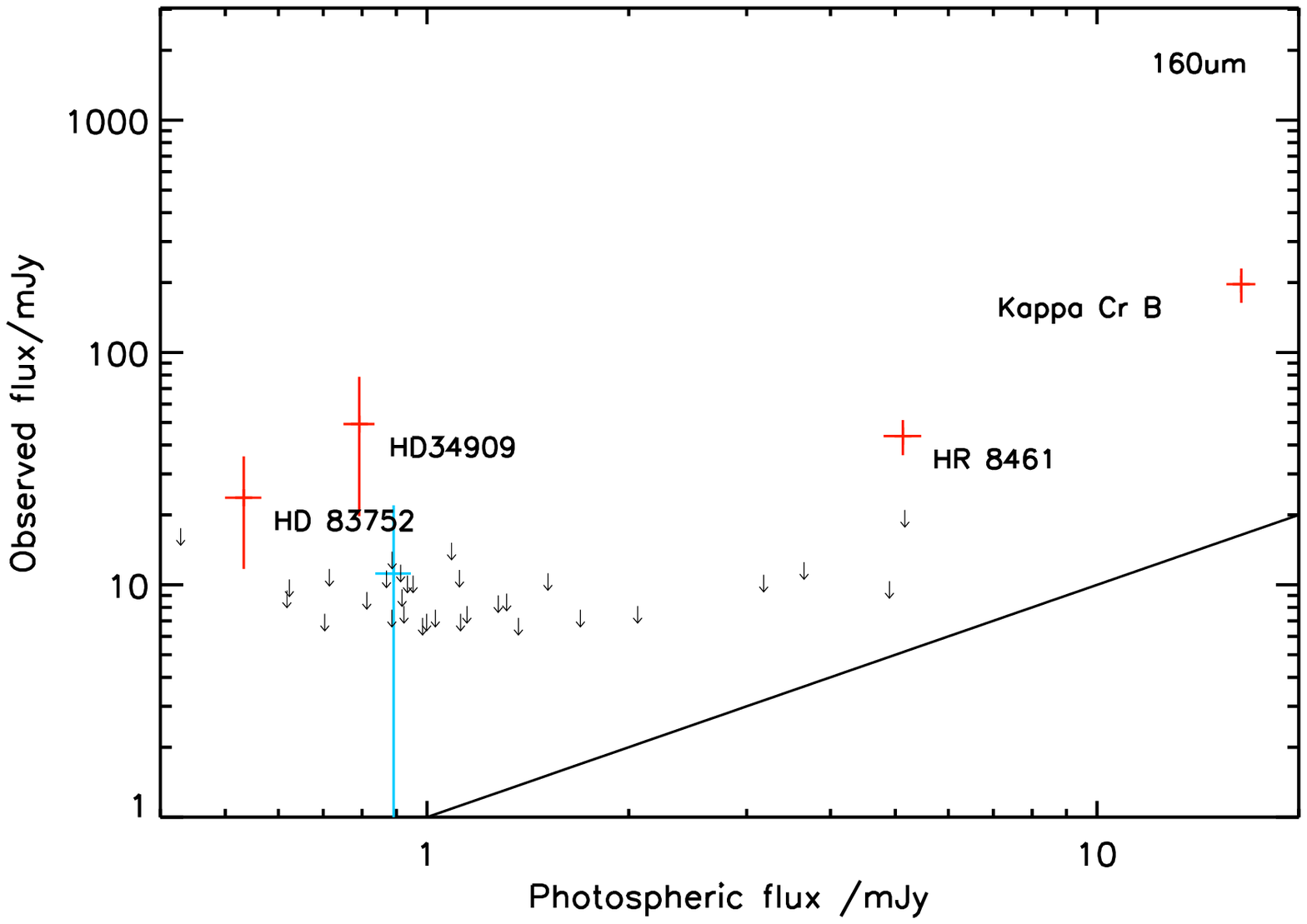}

\caption{The photospheric flux, predicted using stellar models (as discussed in \S\ref{sec:pointsource}) compared to the observed flux at $100\mu$m (top) and $160\mu$m (bottom). The red points show $3\sigma$ detections of an excess, the blue points, $3\sigma$ detections of the system, both with $3\sigma$ error bars, and the black arrows the $3\sigma$ upper limits on the observed flux, when the detection is below $3\sigma$.   } 
\label{fig:fpred_fobs}
\end{figure}

Flux densities for each source are measured by fitting a model PSF (observations of alpha Boo reduced in the same way as the data) at the supposed source location in each image. For non-detections an attempt to make a similar fit is made, but the measured flux is lower than the measured noise. For the sources where the emission was resolved, we do aperture photometry and use a sufficiently large aperture that all disc emission is captured. We assume that the emission is centred on the \Herschel pointing, unless sufficient emission is detected to suggest that the centre is offset from the pointing, in which case the centering is adjusted appropriately.  Uncertainties were calculated using the results from the least squares PSF fitting, and were checked for consistency against the standard deviation of a large number of apertures of sizes $5\arcsec$ and $8\arcsec$ (the sizes for optimal S/N at $100\mu$m and $160\mu$m) and placed randomly near the map centres. These integrated fluxes and uncertainties were compared to predictions for the stellar photosphere, as shown in Fig.~\ref{fig:fpred_fobs}.

Fig.~\ref{fig:fpred_fobs} shows that for the majority of sources, the predicted photospheric flux is too low to have been detected in the \Herschel observations. A significant detection of the source (excess and/or photosphere) was detected for a total of 15 (5) sources at $100\mu$m ($160\mu$m), shown by the blue crosses. Significant emission, $3\sigma$ above the stellar photosphere, shown by the red crosses, was detected for 6 sources at $100\mu$m, namely, \kcrb (HD 142091), HR 8461 (HD 210702), HD 34909, HD 83752, HD 208585 and HD 13496, but only 4 at $160\mu$m,\kcrb (HD 142091), HR 8461 (HD 210702), HD 34909 and HD 83752.

\begin{figure}
\includegraphics[width=0.4\textwidth ]{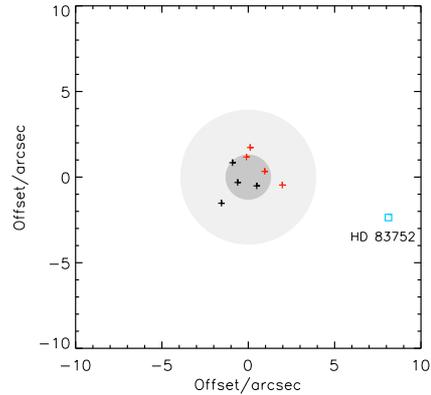}

\caption{The offset of the centre of the observed emission at $100\mu$m, compared to the nominal location of the star. The grey circles show the weighted average absolute pointing error of our \Herschel observations, $1.3\arcsec$ at $1\sigma$ and $3.9\arcsec$ at $3\sigma$. The black points show the debris detections (\kcrbcomma HR 8461, HD 208585 and HD 13496), whilst the red points, show the stars for whom the emission at $100\mu$m was greater than $3\sigma$ above the error on the observations. The blue square shows HD 83752, whose excess emission is suspected of being contaminated with background emission. The emission surrounding HD 34909 is not point-like and, therefore, omitted from this plot}
\label{fig:offset}
\end{figure}

For each of our detections, the offset of the peak from the nominal stellar position was also determined. These are shown in Fig.~\ref{fig:offset}, compared to the pointing accuracy of \Herscheldot All detected sources have small offsets from the nominal \Herschel pointing (close to or less than the $1\sigma$ error), apart from HD 83752, suspected of being contaminated with background emission, as described below. HD 34909 has not been included is in this plot as the emission is not consistent with a point source (see later discussion).

For the sources where the emission was consistent with originating in a debris disc, we used the spectral energy distribution (SED) to obtain an estimate of the disc temperature. In order to determine this, we make the simplest possible assumption; that the dust grains emit like black-bodies. The inefficient emission properties of real grains will reduce the flux at long wavelengths. In order to better model this we have introduced the free parameters $\lambda_0$ and $\beta$ and reduced the black body flux by a factor of $\left (\frac{\lambda}{\lambda_0}\right)^{-\beta}$ at wavelengths longer than $\lambda_0$. $\lambda_0$ and $\beta$ are very poorly constrained, but nonetheless illustrative of the reduced emission anticipated at long wavelengths, that could be relevant for future observations, for example with ALMA.


\subsection{Notes on the 6 sources where an IR excess was detected}
\subsubsection{Kappa Cr B}

\begin{figure}
\includegraphics[width=0.48\textwidth]{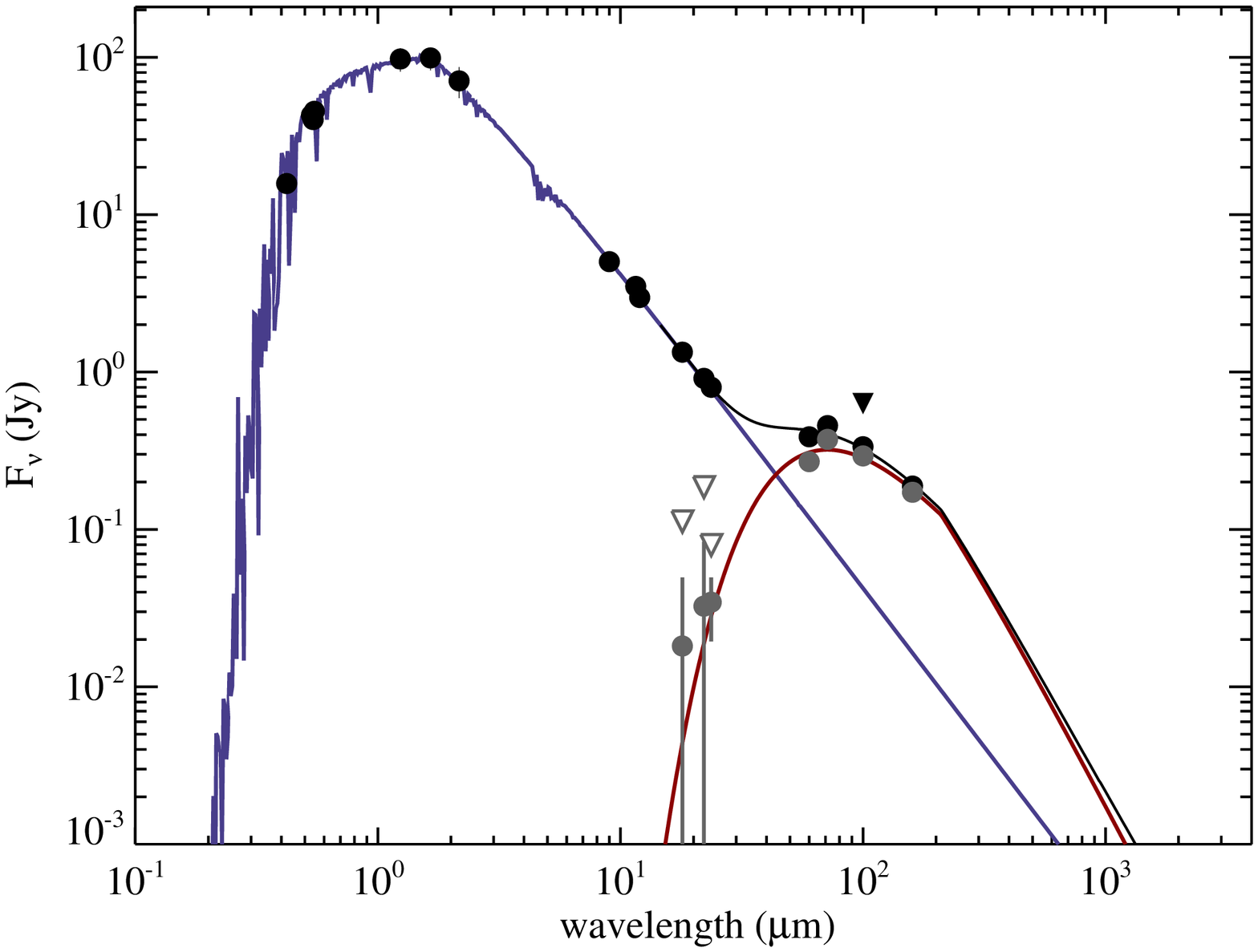}

\caption{SED for \kcrb \citep{bonsor_kcrb}. Photometry is shown as black
dots or black triangles for upper limits. Disc (i.e. photosphere-
subtracted) fluxes and upper limits are shown as grey dots and
open triangles. The stellar spectrum is shown as a blue line and the red line shows a black-body fit to the disc emission, modified at long wavelengths to illustrate the decrease in emission of more realistic grains. The disc has a black-body temperature of $72\pm3$K, which corresponds to a black-body radius of $53\pm5$AU. This is subtly different from \citet{bonsor_kcrb}, where $\lambda_0$ and $\beta$ were free parameters.} 
\label{fig:kappacrb_sed}
\end{figure}

This is the strongest detection in our sample. The debris disc is resolved at both $100\mu$m and $160\mu$m and we refer the reader to \cite{bonsor_kcrb} for detailed modelling of this source. Fig.~\ref{fig:kappacrb_sed} shows a spectral energy distribution (SED) including observations of this source at all wavelengths and a black-body fit to the disc emission. RV observations of \kcrb find evidence for two companions, a close-in, $m\sin I= 2.1M_J$, on a mildly eccentric ($e=0.125 \pm 0.049$) orbit, with a semi-major axis of $2.8\pm0.1$AU \citep{Johnson08} and a second companion, deduced from a trend found in the long-term RV monitoring of this source. AO imaging failed to detect this second companion, thus, if it is on a circular orbit, its semi-major axis lies interior to 70AU (see \cite{bonsor_kcrb} for full details). Modelling of the resolved images combined with the SED information shows that the dusty material is either found in a single wide dust belt, stretching from 20 to 220AU, or two narrow thin dust belts centered on 40 and 165AU. 

\begin{figure*}
\includegraphics[width=1.1\textwidth]{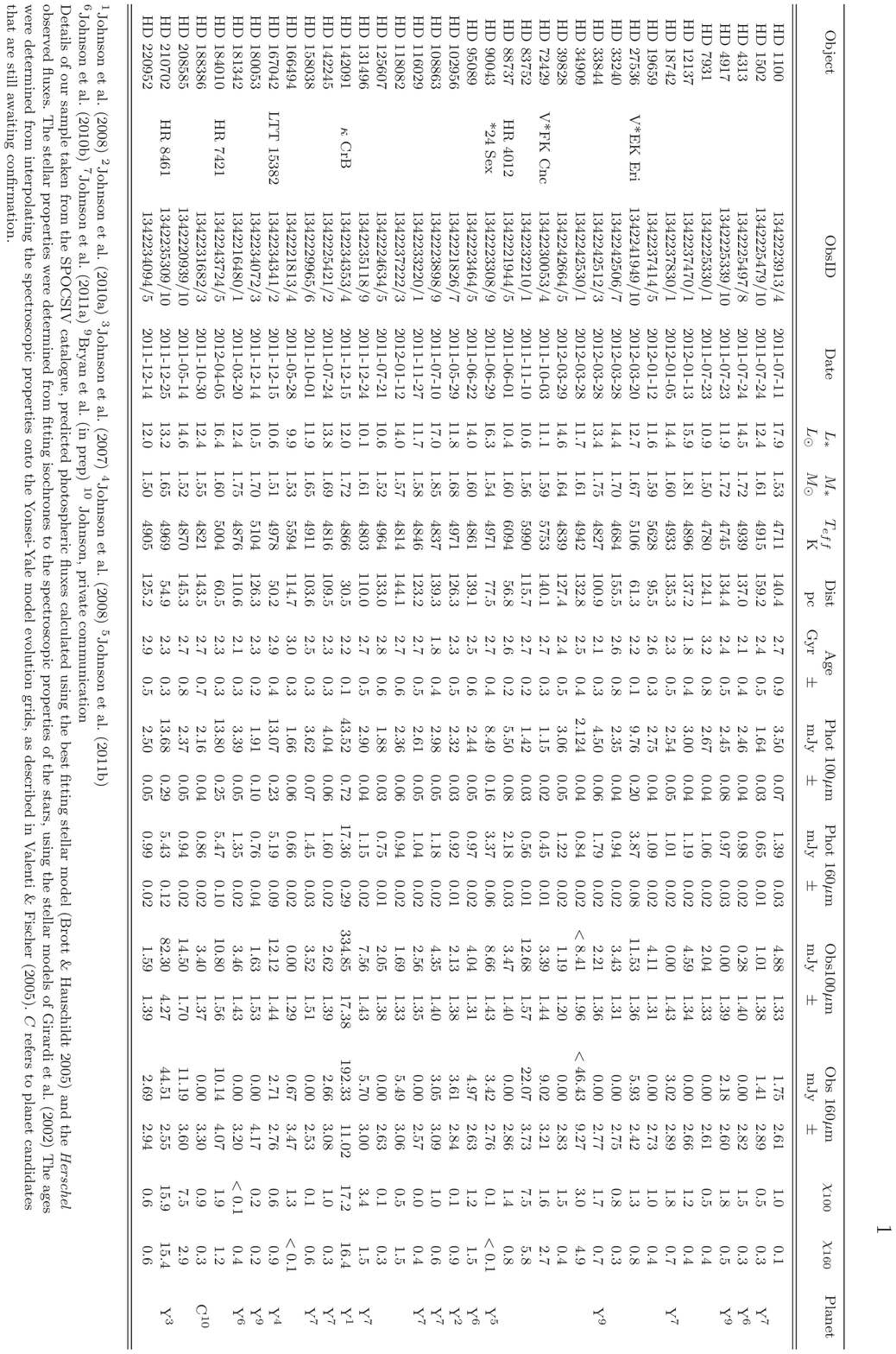}

\end{figure*}

\subsubsection{HR 8461}
\label{sec:HR8461}
\begin{figure}
\includegraphics[width=0.48\textwidth]{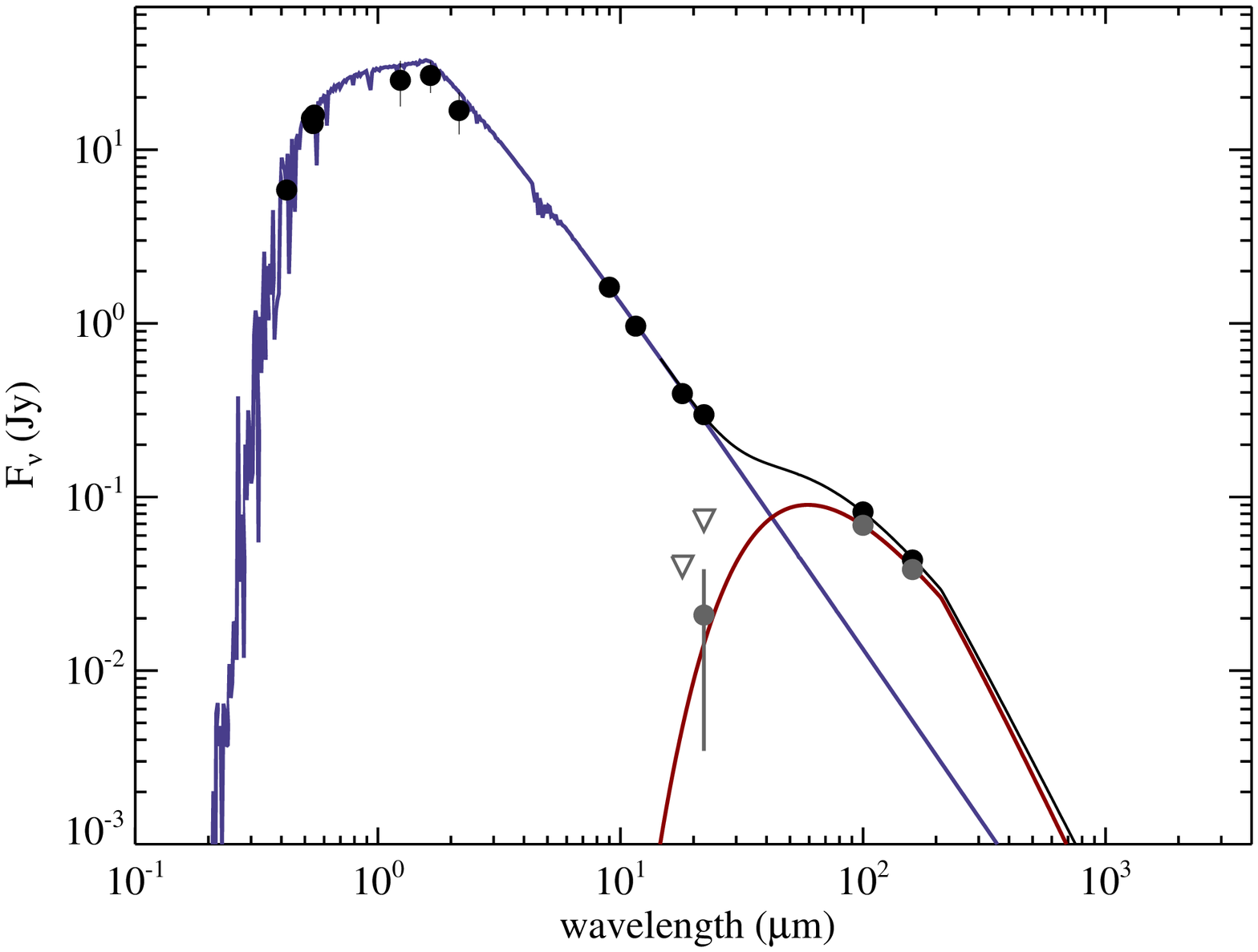}

\caption{SED of HR 8461, plotted in the same manner as Fig.~\ref{fig:kappacrb_sed}, including a WISE measurement at $24\mu$m. The disc has a black-body temperature of $86\pm11$K, which corresponds to a black-body radius of $38\pm10$AU. } 
\label{fig:hr8461_sed}
\end{figure}

\begin{figure}
\includegraphics[width=0.48\textwidth]{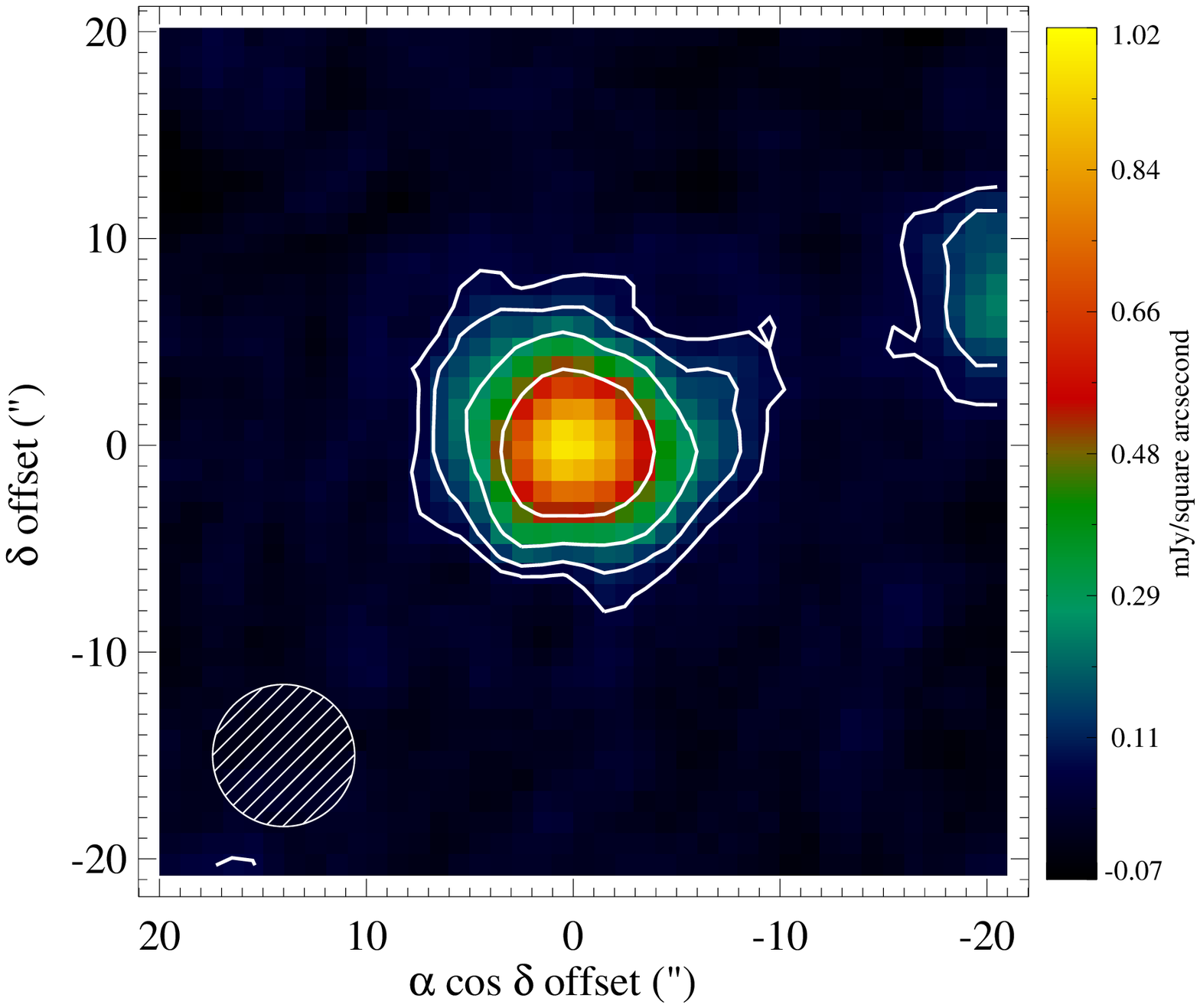}
\includegraphics[width=0.48\textwidth]{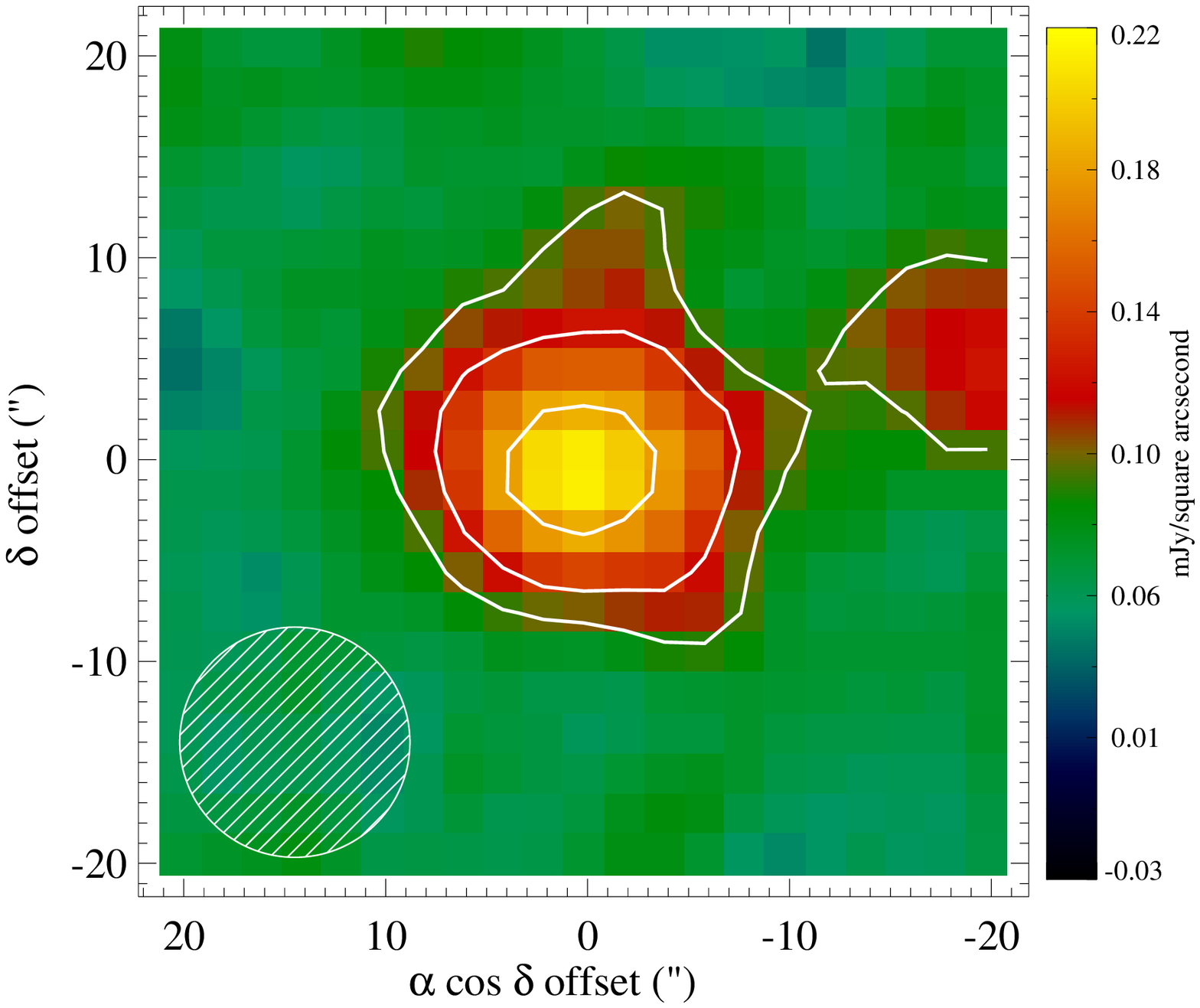}
\includegraphics[width=0.48\textwidth]{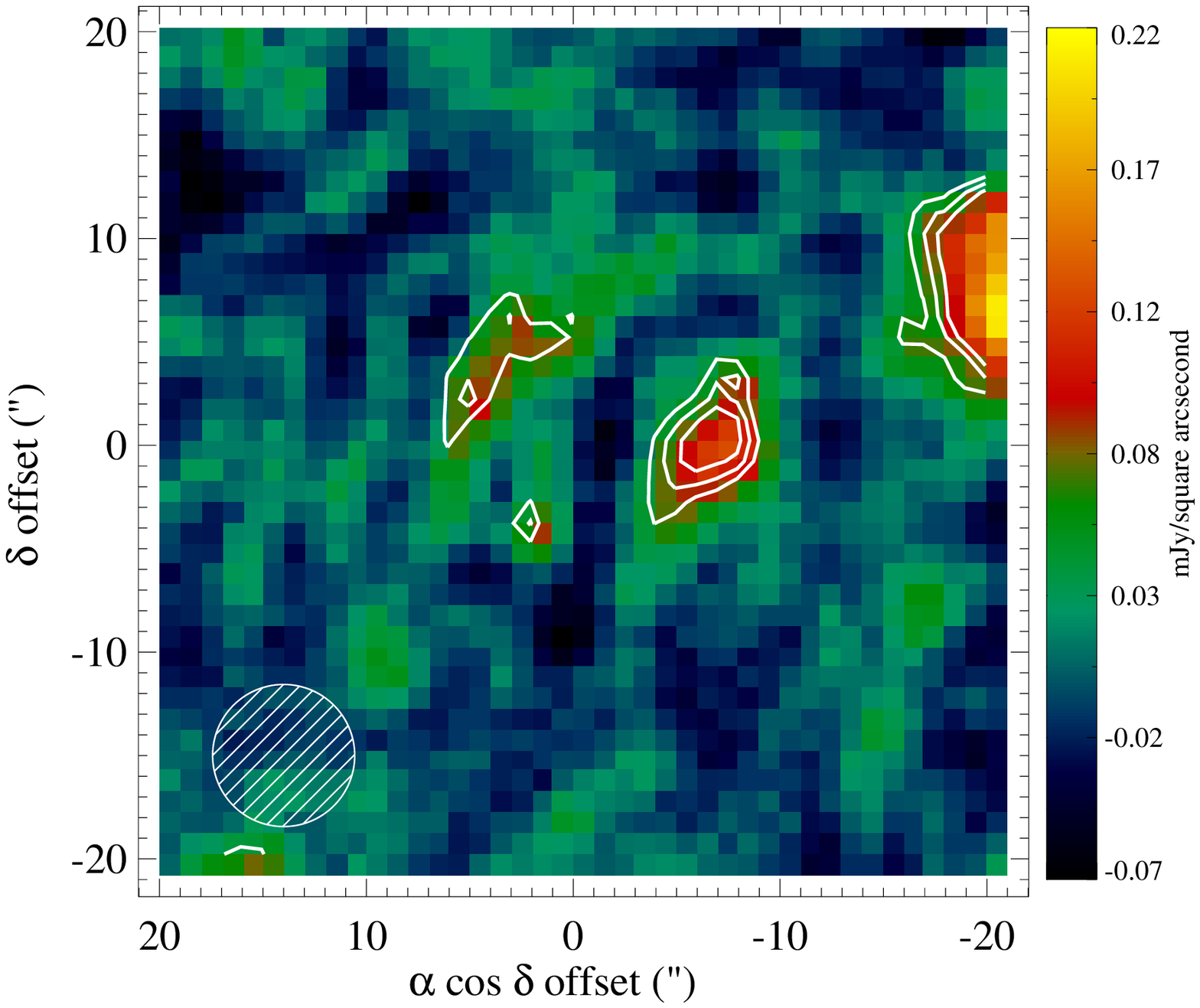}

\caption{The \Herschel images of HR 8461 at $100\mu$m (top) and $160\mu$m (middle), with contours showing detection levels of $3, 6, 12 \sigma$, where $\sigma=2.09\times 10^{-5}$ at $100\mu$m and $\sigma=5.7\times 10^{-5}$ at $160\mu$m. The bottom panel plots the residuals after subtracting a stellar PSF scaled to the peak, with contours at 3,4,5 $\sigma$. This shows that the disc is marginally resolved at $100\mu$m. A similar plot shows that it is not resolved at $160\mu$m. } 
\label{fig:hr8461}
\end{figure}

Radial velocity monitoring found a planet with $M \sin I = 2.0M_J$ in a 341.1 day orbit around HR 8461 \citep{Johnson07}. As can be seen in Fig.~\ref{fig:fpred_fobs}, the emission is clearly significantly above the predicted stellar photosphere at both $100\mu$m ($16\sigma$) and $160\mu$m ($15\sigma$). The \Herschel data points are plotted alongside the stellar spectrum in Fig.~\ref{fig:hr8461_sed}. A black-body fit to the data yields a black-body temperature of $86\pm11$K, which would correspond to a disc radius of $37.9\pm9.7$AU if the dust acts like a black body.  Fig.~\ref{fig:hr8461} shows the \Herschel images of this source at both $100\mu$m and $160\mu$m. The disc is marginally resolved at $100\mu$m, as shown by star-subtracted image in the bottom panel of Fig.~\ref{fig:hr8461}. Fitting a simple ring to the image finds a disc radius of roughly 130AU, somewhat larger than the black-body radius, but not inconsistent with models that suggest that grains with realistic emission properties can be significantly hotter than black-bodies \citep{bonsor10, Kains11, Booth2012}. The position angle and inclination of the disc in this fit are poorly constrained.

\subsubsection{HD 208585}

\begin{figure}
\includegraphics[width=0.48\textwidth]{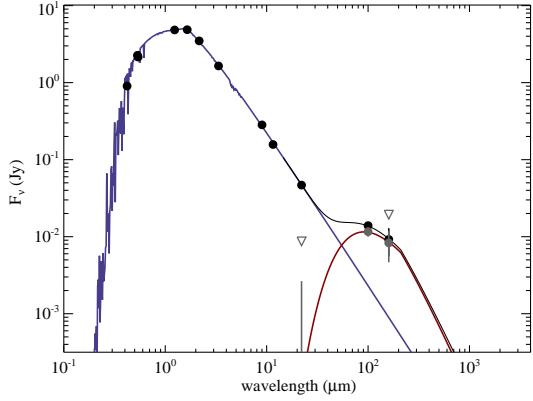}

\caption{SED of HD 208585, plotted in the same manner as Fig.~\ref{fig:kappacrb_sed}. The disc emission is not well constrained, but we fit a black-body temperature of $56^{+65}_{-30}$K, which corresponds to a black-body radius of $94^{+440}_{-20}$AU. } 
\label{fig:hd208585}
\end{figure}

A 100$\mu$m excess is clearly detected (7.1 $\sigma$) for HD 208585, and a marginal excess at $160\mu$m, with a significance of $2.8\sigma$. The emission is not resolved at either wavelength. In Fig.~\ref{fig:hd208585} we plot the \Herschel data points, alongside the full stellar spectrum.  The temperature is poorly constrained due to the marginal detection at 160um; the blackbody fit yields 56K with a 3$\sigma$ range of 25-120K. The wide range of disc temperatures, and the fact that the disc temperature and normalisation are correlated, mean that the uncertainties are asymmetric about 56K. They were estimated using a grid in temperature vs. normalisation space that calculates the deviation from the best fit $\chi^2$ at each point.

This source has been searched for radial velocity companions, with a total of 11 RV measurements, over a time period spanning 5 years, but currently there have been no detections of any companions.

\subsubsection{HD 131496}

\begin{figure}
\includegraphics[width=0.48\textwidth]{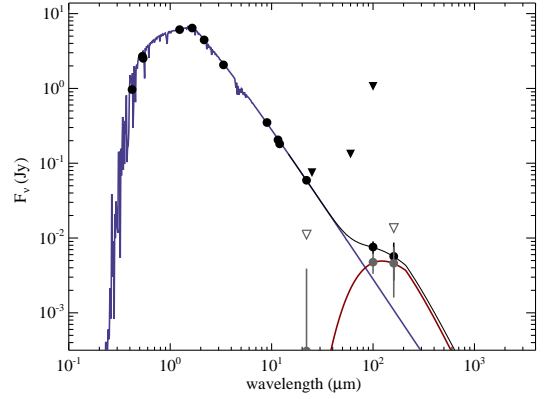}

\caption{SED of HD 131496, plotted in the same manner as Fig.~\ref{fig:kappacrb_sed}, but with only a single data point the disc temperature is poorly constrained.  } 
\label{fig:HD131496}
\end{figure}

A $2.2M_J$ planet was detected orbiting HD 131496 by \cite{Johnson2011}. Excess emission was found at this source at a level of $3.4\sigma$ at $100\mu$m, but not detected at $160\mu$m ($1.5\sigma$ excess). As for HD 208585, the disc temperature is not well constrained, with a best fit of 42K and a 3$\sigma$ range of 10-370K. 
 Fig.~\ref{fig:HD131496} shows the SED for HD 131496.

\begin{figure*}
\includegraphics[width=0.48\textwidth]{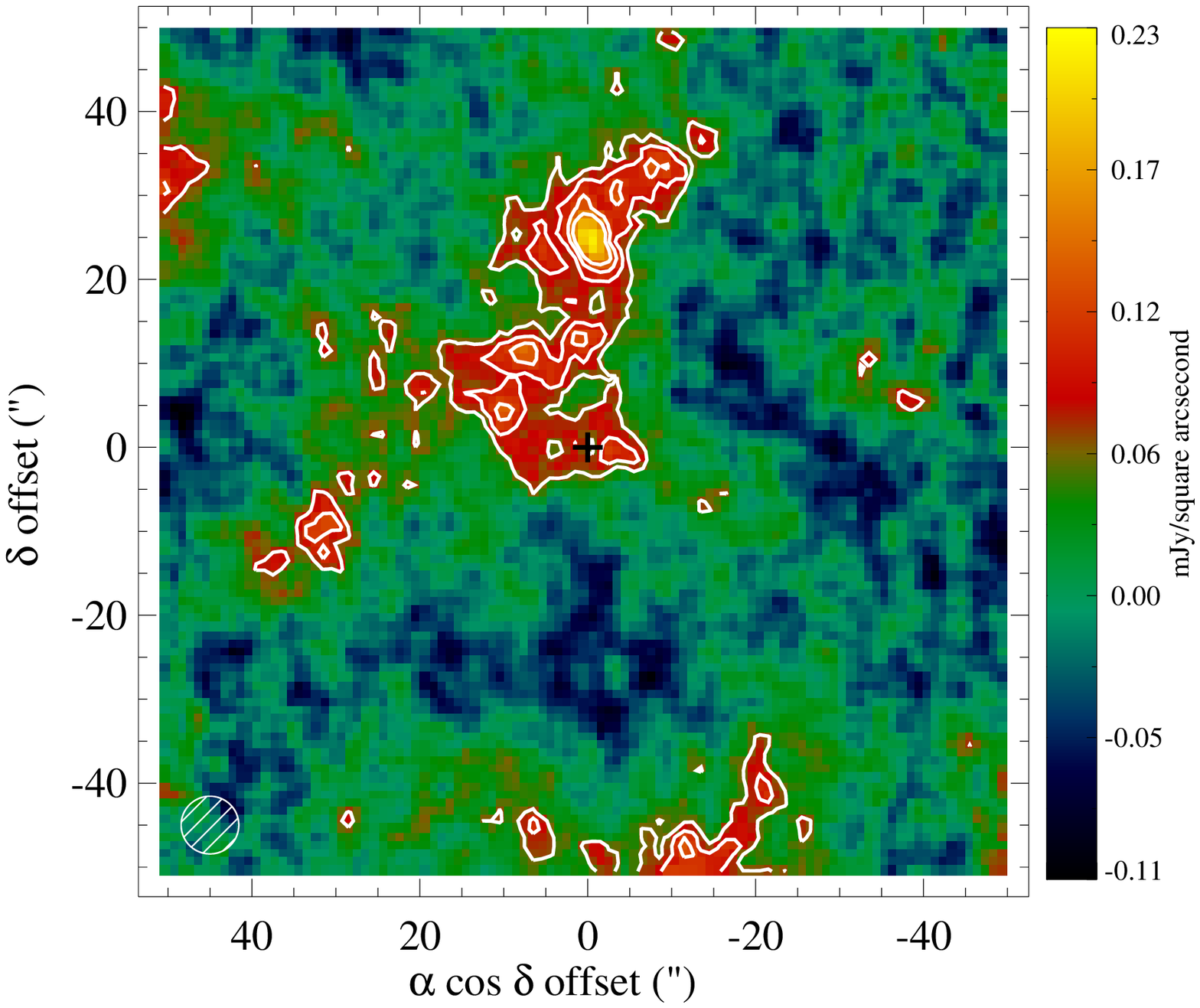}
\includegraphics[width=0.48\textwidth]{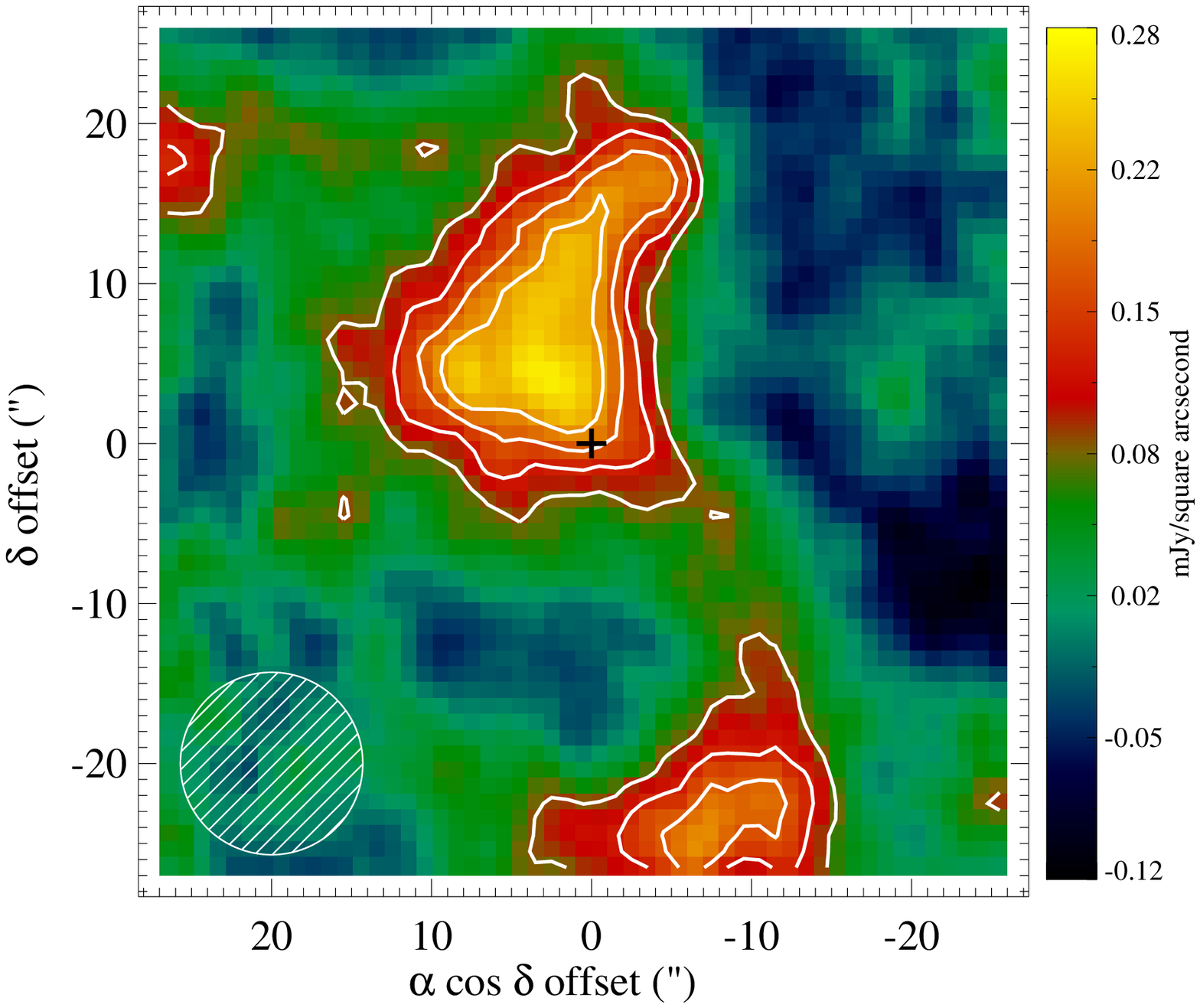}
\caption{The \Herschel images of HD 34909 at $100$ (left) and $160\mu$m (right) showing clear contamination from the dusty, star-forming region at a similar location on the sky.  The cross shows the nominal location of the star and the contours show 2,3,4,$5\sigma$ detection levels, where $\sigma=2.9\times10^{-5}$ at $100\mu$m and $\sigma=1.7\times10^{-4}$ at $160\mu$m.} 
\label{fig:HD34909}
\end{figure*}

\subsubsection{HD 83752}
The emission observed close to this object is likely to be from a background source. The predicted stellar photosphere is significantly below the \Herschel detection limits at both wavelengths and the emission detected in the field of view of this object is offset from the nominal stellar position by $8.5\arcsec$, larger than the $3.1\arcsec$, $3\sigma$ absolute pointing error of the \Herschel observations for this target, as can be seen in Fig.~\ref{fig:offset}. In addition to this the observed emission increases from $100\mu$m to $160\mu$m, by a factor of 2, a feature generally reminiscent of the emission from cold, distant galaxies. The best fit black-body temperature would need to be less than 30K, equivalent to black-body grains situated further than 350AU from this $10.7L_\odot$ star. Although there have been previous detections of cold debris discs with \Herschel \citep{Krivov2013}, in this case a diameter of 700AU corresponds to 6" at 116pc, and if this really were a debris disc of black body size or larger, it would appear extended in our \Herschel observations, assuming that the dusty material is not found in a clump at this radial distance. We, therefore, conclude that this emission likely originates in a background galaxy, not a debris disc orbiting HD 83752.


\subsubsection{HD 34909}
The \Herschel image (Fig.~\ref{fig:HD34909}) of this source reveals extended emission filling a significant proportion of the field of view at both $100\mu$m and $160\mu$m. HD 34909 is situated in a very dusty region of the sky, at $23^\circ$ from the galactic plane, close to a star-forming region in Orion. The emission observed is likely to result from dust in this molecular cloud and it is very difficult to assess whether there is additional emission from the subgiant, or a potential debris disc orbiting HD 34909. We place an upper limit on the flux from a disc that could be hidden in this system (see Table 1.).

\subsection{Background galaxy contamination in our sample}
\label{sec:background}
Contamination of observations at the \Herschel wavelengths by background sources is common. The level of such contamination in PACS data is characterised by \cite{Sibthorpe2013}. The contamination in our observations can be assessed in comparison with such results. If we consider the weighted average $3\sigma$ absolute pointing error of \Herschel PACS in our observations of $3.94\arcsec$ and the average $3\sigma$ error on our observations of $\pm5.7$mJy at $100\mu$m, according to the results of \cite{Sibthorpe2013}, there is a 1.5\% chance of confusion by one or more background sources. Given that we have observed 36 stars, this means that the probability of no stars being contaminated is 58\%,
whilst the probability of one star being contaminated by a background detection is 31.8\% 
and of two stars being contaminated is 4.8\%. This is critically important in our assessment of our detection statistics as it means that the  chances of one of our detections resulting from a background object is not insigificant (31.8\%), although there is only a slim chance of more than one of our detections being contaminated in this manner (4.8\%). We should, however, note that for some of our detections there is good evidence to suggest that we are observing debris discs rather than galaxies, for example, the extended nature of \kcrb and HR 8761.

We note that both HD 83752 and HD 34909 should be excluded from the above analysis. HD 34909 is a separate case as it is located in a region of the sky known to be contaminated by the dusty emission from star formation. HD 83752 has an offset of $8.5\arcsec$ from the nominal \Herschel pointing, more than $3\sigma$ from the nominal \Herschel pointing.  There is a 96\% chance of one of our stars having a detectable background source within $10\arcsec$ of the \Herschel pointing for our stars. Thus, our detection of emission close to HD 83752 is consistent with our statistical analysis.

\subsection{Summary}

Excess emission, consistent with a debris disc, was found for 4/36 (2/36) of the subgiants in our sample, namely \kcrbcomma HR 8461, HD 208585 and HD 131496 (\kcrb and HR 8461) at $100\mu$m (160$\mu$m). Three of the detections are planet-hosting stars (\kcrbcomma HR 8461 and HD 131496), whilst HD 208585 has no current planet detections. This leaves us with 3/19 planet hosts that have debris and 1/17 control (non-planet host) stars with debris.

\section{Comparison with observations of debris discs orbiting main-sequence stars }
\label{sec:model}

Our 4 debris disc detections for subgiants provide a tiny sample compared to the wealth of observations already obtained for main-sequence stars. It is, nonetheless, interesting to assess whether our sample shows similar trends to those observed in the population of debris discs orbiting main-sequence stars. A prime example being the decrease in brightness of the discs with age of the star. Subgiants, being older than their main-sequence counter parts (at least for similar mass stars), provide crucial information regarding the continuation of this trend to later times.

A simple toy model developed in \cite{Wyatt07hot, wyatt07} provides a reasonable description of the time evolution of the disc luminosity, by considering each disc to have a flat, then 1/age dependence, which is a reasonable approximation to models that consider the evolution of the size distribution in more detail \citep[e.g.][]{lohne, gaspar2013}. This fits the observational data for both debris discs orbiting main-sequence A stars \citep{wyatt07} and FGK stars \citep{Kains11}. Here, we compute a model population, based on the work of \cite{wyatt07}, but applied to the specific properties, \eg age, luminosity and distance, of our sample of subgiants.  This model, therefore, takes into account any differences between our subgiants and main-sequence stars, including evolution in luminosity or the effects of radiation pressure that remove the smallest grains in the disc. Initially, we focus on the fate of debris discs observed around main-sequence A stars, using the model parameters derived in \cite{wyatt07}, as isochrone fitting using the stellar models of \cite{Girardi2002} suggests that our subgiants evolved from similar mass stars \citep{Johnson10_subgiant, Johnson07, Johnson08, Johnson10, Johnson2011,Johnson2011_24Sex} .


\begin{figure}
\includegraphics[width=0.48\textwidth]{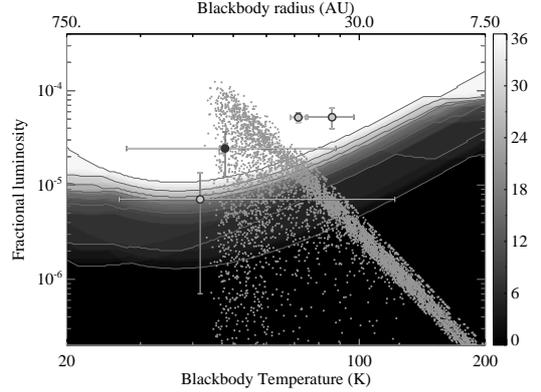}

\caption{ The cumulative detection limits for discs orbiting all of the stars in our sample, as a function of the disc black-body temperature and fractional luminosity (ratio of the total disc luminosity to the total stellar luminosity), with $1\sigma$ error bars. The equivalent black-body radius for a disc orbiting a $15L_\odot$ star is plotted on the top axis for reference. The discs that were detected in our sample are plotted by the large circles, empty circles for the planet-hosts and filled circles for the non-planet hosts, whilst the small circles show the model population, as described in \S\ref{sec:model}. It should be noted that there is an artificial upper (lower) limit in radius (black-body temperature), which is imposed by the model population based on the work of \citet{wyatt07}. }
\label{fig:f_r}
\end{figure}

This model population is plotted in terms of its fractional luminosity and temperature in Fig.~\ref{fig:f_r}. The detections from our observations are shown (circles) for comparison.  The top axis shows the equivalent black-body temperature for a disc orbiting a $15L_\odot$ star for reference, although, we point out here that the real radii of detected discs may differ from their black-body radii. This model population shows a clear upper limit in fractional luminosity as a function of temperature (or radius), with no discs inhabiting the top right corner of this plot. As collisions grind down the material in the debris discs, with faster evolution for closer in discs, there is a maximum mass that can survive in steady-state as a function of time and disc radius. The similar ages of our stars (see Table 1.) means that this maximum mass, which translates into a maximum fractional luminosity, is predominantly a function of the disc radius, and can be clearly seen in the upper envelope of the model population on Fig.~\ref{fig:f_r}. Two of our detections (\kcrb and HR 8461) lie above this maximum. This does not mean that these discs need be unphysically massive, since the model population is derived with the parameters of an average disc, whereas some discs would be expected to have maximum disc masses that can lie an order of magnitude above that of the average disc, and it is such discs that would be most easily detected. However, an alternative explanation could be that some individual sources do not fit within the context of this simple model. Notably, the resolved images of \kcrb indicate that the emission results from either a wide belt, or two distinct belts \citep{bonsor_kcrb}, whereas the model assumed all discs to be described by a single narrow ring.

\begin{figure}
\includegraphics[width=0.48\textwidth]{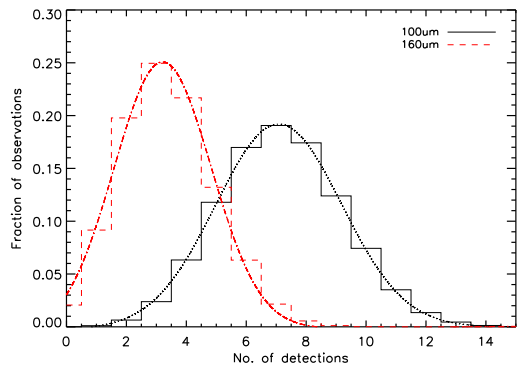}

\caption{A histogram to show the number of detections found after repeated observations of the model population at $100\mu$m (black solid line) and $160\mu$m (red dashed line). The mean detection rate was found to be 7/36 at $100\mu$m and 3.4/36 at $160\mu$m. A Gaussian was fitted to distribution, with width $\sigma=2$ ($100\mu$m, black dotted line) and $\sigma=1.6$ ($160\mu$m, red dash-dotted line).}
\label{fig:det_prob}
\end{figure}

Only a small fraction of the model population could have been detected, due to the limited sensitivity of our observations. We characterised this, by calculating whether a simple, thin, black-body disc at each temperature, with each fractional luminosity, would have been detected orbiting each star in our survey. The grey lines on Fig.~\ref{fig:f_r} show the fraction of this model population that would have been detected in our \Herschel observations. The white (black) shaded area shows the parameter space where debris discs would have be detected around all (no) stars in our survey. In order to compare the model population with our observations, we calculate the detection rate that would be obtained if this model population were observed with \Herschel and the observation limits of our survey. This is done by assigning each star a disc from the model population and comparing the emission from the disc to the sensitivity of the \Herschel observations obtained for that source, thus, determining whether or not the disc would have been detected. We note here that are survey detects debris discs more readily around nearby stars. Repetition of this process leads to a range of detection rates, as shown in Fig.~\ref{fig:det_prob}. The mean detection rate obtained  is 7/36 (3.4/36) at 100 (160)$\mu$m, which is higher than the detection rates found in our observations of 4/36 (2/36). However, further analysis finds that the probability of detecting only 4/36(2/36) debris discs, given this distribution is within $2\sigma$ ($1\sigma$) of the mean detection rate at $100\mu$m ($160\mu$m). In other words our detection rate is on the low side, but not inconsistent with being derived from the model population.

Although not required to explain the observations, one potential issue with the comparison between the model population and our observations should be noted here. The model of Fig.~\ref{fig:f_r} for the debris discs orbiting our subgiants assumed that these are descended from a main sequence population with disc parameters (e.g., planetesimal strength, size and orbital eccentricities) that were optimised to explain observations of the evolution of debris discs around main sequence A stars \citep{su06, rieke05, wyatt07}, with spectral types B8V to A9V. However, the disc parameters required to explain the population of main sequence discs around Sun-like stars is different \citep{Kains11}, as is the main sequence lifetime. Thus, if our stars evolved from a different distribution of spectral types on the main-sequence, our predicted detection statistics could vary significantly. This detail is particularly relevant given the current discussion in the literature regarding the masses of planet-hosting subgiants, including those used in this survey \citep{Lloyd2011, Lloyd2013, Schlaufman2013}. Had the rate we observed been significantly different to that predicted from the model, this could have been interpreted in terms of the properties of the main sequence progenitors. As it is, there is no evidence to exclude these subgiants' discs being evolved versions of the population seen around main sequence A stars, particularly when it is noted that additional factors, such as dynamical instabilities, tend to reduce the fraction of stars with detectable debris around older stars.


To summarise, our observations do not currently provide any evidence against a very simple model for the collisional evolution of debris discs from the main-sequence to the subgiant branch, nor do they provide further evidence regarding the main-sequence origin of our subgiant sample. As further detections of debris discs orbiting subgiants are obtained this analysis can be extended and the new data used to repeat the analysis of \cite{wyatt07} or \cite{Kains11} including a further time bin for stars on the subgiant branch, thus, further constraining our understanding of the collisional evolution of debris discs.

\section{Conclusions}
\label{sec:conclusions}
We have observed 36 subgiants with \Herschel PACS at $100\mu$m and $160\mu$m to search for the presence of excess emission from debris discs. All our stars have been searched for the presence of radial velocity companions as part of the Johnson et al. program at the Keck/Lick observatories, with close-in, planetary companions detected for 20/36 stars. We detected excess emission, thought to be from debris discs, around 4/36 (2/36) of our sample at $100\mu$m ($160\mu$m). Observations at \Herschel wavelengths are frequently contaminated by emission from background galaxies. Whilst we acknowledge that there is a 30\% chance of emission from a background galaxy around one of our sources mimicing a debris disc \citep{Sibthorpe2013}, we note that there is a $<5\%$ chance of contamination for more than one of our stars. Our detections form 3/19 of the planet-host sample and 1/17 of the control sample. Such a small number of detections provide no evidence that the detection rate for debris discs around stars with planets is different to that around stars without planets. These observations provide an important constraint for debris disc models, informing us of the fraction of stars with detectable discs at the end of the main-sequence, which should be included in future models of debris disc evolution. These detections illustrate that large quantities of dusty material can survive the star's main-sequence evolution, providing a potential link with observations of pollution and dust around white dwarfs.

\section{Acknowledgements} 

AB acknowledges the support of the ANR-2010 BLAN-0505-01 (EXOZODI). MCW and GK acknowledge the support of the European Union through ERC grant number 279973. We thank Marta Bryan for useful discussions that improved the quality of this manuscript.

\bibliographystyle{mn}

\bibliography{ref}

\end{document}